# Improving the Cybersecurity of Critical National Infrastructure using Modelling and Simulation

Uchenna D Ani[1], Jeremy D McK Watson[2], Nilufer Tuptuk[3], Steve Hailes[4], Madeline Carr[5], Carsten Maple[6].

[1]School of Computing and Mathematics, Keele University, UK
[2]Department of Science Technology Engineering and Public Policy, University College London
[3]Department of Security and Crime Science, University College London
[4,5]Department of Computer Science, University College London
[6]Cybersecurity Centre, WMG, University of Warwick, UK

This policy note aims to raise awareness of the importance of adopting cybersecurity modelling and simulation techniques that embody integrated social and technical factors. It is based on research and synthesis following a state-of-the-art literature survey and engagement workshop with critical infrastructure stakeholders hosted by the Department for Transport (DfT) in February 2019, and a desk-based study in the ongoing (2020-2022) PETRAS Modelling for Socio-technical Security (MASS) project. Participants from academia, business and industrial sectors, and government came together to discuss the effectiveness of modelling and simulation to support the protection of modern Critical Infrastructure Systems. The discussion also covered how government effort, through policy interventions, can support the National Cyber Security Strategy 2016-2021, and beyond. Although the focus of the initial workshop was the transport sector, the insights drawn and recommendations made can be applied to other critical national infrastructure sectors such as Energy, Water, Defence, Chemicals, and Food.

**This policy briefing is targeted at government departments and agencies with management and regulatory roles linked to critical infrastructure sectors listed above. The advice may be particularly relevant to DfT, Cabinet Office, BEIS, Home Office, DCMS, DEFRA, and GO-Science.**

## *Summary:*

The UK's Critical National Infrastructure (CNI) is critically dependent on digital technologies that provide communications, monitoring, control, and decision-support functionalities. Digital technologies are progressively enhancing efficiency, reliability, and availability of infrastructure, and enabling new benefits not previously available. These benefits can introduce vulnerabilities through the connectivity enabled by the digital systems, thus, making it easier for would-be attackers, who frequently use socio-technical approaches, exploiting 'humans in the loop' to break in and sabotage an organisation. Therefore, policies and strategies that minimise and manage risks must include an understanding of operator and corporate behaviours, as well as technical elements and the interfaces between them and humans.

Gaps exist in organisational understanding at several levels within the least aware critical infrastructure organisations. At the most basic, operators may not understand the degrees of systems vulnerability or the types and subtlety of attacks. More aware infrastructure organisations deal with cybersecurity at a purely technical level but exclude behavioural and other social factors. Best in class approaches use a holistic socio-technical viewpoint and are moving towards modelling and simulation (M&S) as a means of testing and assuring cybersecurity measures for securing modern critical infrastructure systems. Better security via socio-technical security M&S can be achieved if backed by government effort, including appropriate policy interventions. Government, through its departments and agencies, can contribute by sign-posting and shaping the decision-making environment concerning cybersecurity M&S approaches and tools, showing how they can contribute to enhancing security in Modern Critical Infrastructure Systems.

*Key recommendations for policymakers, to:*

1. **Encourage wider awareness and adoption of a socio-technical approach to security, from system M&S to implementation in CNIs.**
2. **Promote the use of critical infrastructure security open-source M&S approaches – tools and techniques.**
3. **Establish policies and platforms that encourage evaluating the credibility of security M&S approaches – tools and techniques.**
4. **Create and maintain governance for security M&S.**
5. **Support open, continuous cross-sector collaboration and knowledge exchange.**
6. **Promote team-based collaborative development of security M&S tools amongst academic/research, industry, and government organisations.**
7. **Incentivise and/or reward the dissemination of security M&S research outcomes**.

## Modelling and Simulation in Critical National Infrastructure Protection

Critical National Infrastructure (CNI) refers to assets, facilities, systems, networks or processes, and the essential workers that operate and facilitate them, the loss or compromise of which could result in 'major detrimental impact on the availability, integrity or delivery of essential societal services'[1]. CNI systems are fundamental and critical to the maintenance of vital societal functions, health, safety, security, economic and social well-being of people[2]. They are critical because interrupting, disrupting or destroying their functions would have serious physical, economic, and social consequences and impacts[3]. However, while security solutions require auditing and assurance before deployment to CNIs, it is often not feasible to test the security characteristics of live operational CNI environments due to the disruptive effect that could occur [4]. To overcome these concerns, modelling and simulation (M&S) provides a viable alternative, replicating CNI environments so cyber security challenges can be explored and addressed [5]. Security M&S is the process of creating a representation of a security situation to show or test how to attack or defend systems and/or users by imitating security-related behaviours, conditions, or other activities of the system or users in relation to cyber security threats, vulnerabilities, or controls. This can support early and better analysis of potential security threats, recognising and appropriately managing associated cyber security risks [6]. Moreover, in an operational context, security M&S of CNI systems can help visualise, manage, control, predict and optimise complex security situations by comparing modelled states with real-world information.

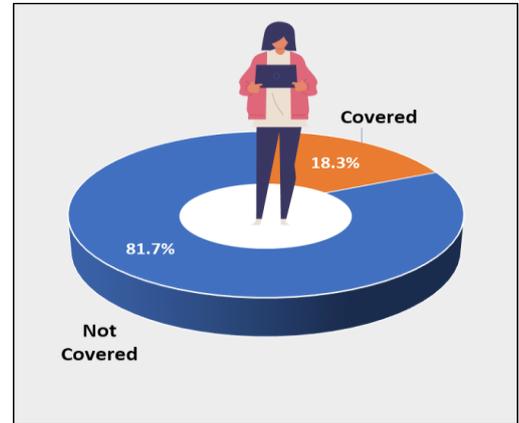

*Figure 1: Distribution of CIP Approaches that Support the Development of Security Policies & Regulations [9,10]*

In February 2018, the UK Government Office for Science and the Council for Science and Technology published a joint Blackett Review on *'Computational modelling: technological futures'*[7]. This report highlighted the importance of M&S in helping to drive performance improvement of products and services, achieve productivity and efficiency gains, and create new innovative smart products and services in CNI sectors such as transportation, energy generation and distribution, water, manufacturing, and healthcare, etc. This category of systems is a prime target of nation-states and terrorist attacks, but now it has become an even larger attack surface as emerging technologies like the Internet of Things (IoT) including Commercial Off-The-Shelf (COTS) hardware and software, and Edge computing devices are being integrated into CNI. These new attack surfaces have potentially serious consequences that cannot be ignored[8].

> *M&S is an unrealised opportunity for CNI protection, that can include new technologies and practices which reflect the fast-moving dynamics of evolving security needs*

There is a significant awareness gap in the critical infrastructure community regarding state-of-the-art M&S and its role in informing the scale and extent of security weaknesses, particularly where emerging technologies such as IoT are applied. Our research [9,10] shown in Figure 1. indicates that most Critical Infrastructure Protection (CIP) approaches lack contexts that support developing security policies and regulations. M&S is an unrealised opportunity for CNI protection and that can include new technologies and practices, which reflect the fast-moving dynamics of evolving security needs[3]. Along with the benefits that new technology trends bring to CNI operations and services, emerging technologies are also expanding attack surfaces in complex infrastructure systems, and increasing the potential frequency and sophistication of cyber-attacks. Thus, the growing threat landscape is driving increasing concerns about how to protect these systems given the vital positions they occupy in maintaining the national socio-economic ecosystems[11], especially because of the potential negative and sometimes dangerous impacts that can occur when unassured security measures and controls are applied in these systems.

## Current Government Initiatives

A central element of the UK's 21st-century global identity is the goal of being the *'safest place to live and do business online'*[12], but this can only happen and be sustained if CNIs function properly, are robust and resilient. This suggests that initiatives are required which can help assure the safety, security, and resilience of CNI. M&S can help to achieve this. Security M&S creates a normalised view of a system or organisation's security situation via models, allowing security-related conditions, behaviours, activities, or processes to be explored using known information on security risk factors (threats, vulnerabilities, attack likelihoods, and impacts), and the effectiveness of available security controls. M&S offers focussed methods for analysing security-related perturbations in such systems and enables evaluation of interdependency and cascading effects across networks of infrastructures, based on their interactions[13]. The UK recognises the value of M&S as an aid to understanding and predicting system behaviours and its role in supporting appropriate decision-making in both the public and private sectors[7]. M&S can help with managing the growing system complexities of modern infrastructures that support human society[14]. These rarely operate in isolation, but are interconnected and generate huge volumes of data that need to be modelled and analysed if useful insights are to be drawn to assure appropriate levels of functionality and security[15], resilience, and safety.

*Security M&S of CNI systems can help visualise, predict, optimise, manage, and control complex security situations by comparing modelled states with real-world evidence*

Efforts are underway in academic, government, and industrial communities to explore practical M&S pathways and develop techniques and tools that advance the protection of critical industrial systems[3,6]. The ongoing National Digital Twin programme (NDTp) run by the Centre for Digital Built Britain (CDBB) launched by HM Treasury in July 2018 is a good example of an initiative exploring M&S approaches to promote better outcomes including security for the built environment[16]. In addition to supporting the development of technologies and methodologies for security M&S, policymakers can intervene in other ways to promote the protection of CNIs[9,10]. These include setting best practice guidelines and procurement rules for public sector projects. Politicians and policy stakeholders need to understand more, and to act to establish and support modelling and simulation-based approaches – including raising awareness, providing best practice guidelines and training at all levels – that can strengthen the capability of critical infrastructure owners and operators to guard against cyber-attacks. This highlights one of the recommendations of the *'Data for the Public Good'* Report[17] by the National Infrastructure Commission in 2017, which led to the role of the NDTp in streamlining the management (including the security) of physical infrastructure. This briefing is intended to highlight the importance of M&S approaches in supporting efficient measures for the safety, security, and resilience of modern CNI systems, and recommending policy pathways to achieve these objectives.

**Recommendations**:
The following recommendations suggest how government can further contribute to shaping and signposting the cybersecurity environment, and how decision-making concerning M&S approaches and tools can enhance security in modern Critical Infrastructure Systems.

1. **Encourage wider awareness and adoption of a socio-technical approach to security, from system M&S to implementation in CNIs**

Security issues today are largely socio-technical, however, few CNI sectors in the UK appear to be exploring a socio-technical approach to security M&S and analysis. IT and Telecommunications, Government facilities, and Financial services seem to be in the lead as shown in Figure 2. There appears to be low awareness and understanding of the value of socio-technical security approaches in addressing the evolving threats to CNI safety, security, resilience. This is evidenced by weak adoption in highly critical sectors like energy, transport, and water, and suggest the need for urgent action. ***Government, through its departments and agencies like the National Cyber Security Centre (NCSC) and Centre for the Protection of National Infrastructure (CPNI), should increase awareness of, and further encourage adoption of socio-technical security thinking and approaches in practical CNI system development, also audit and project sign-off.*** This is especially crucial for CNI because of the significant consequences of their disruption or destruction. Government regulatory agencies can increase collaboration with security engineering organisations, research groups/institutes, and forums similar to their current partnership with 'The Research Institute for Sociotechnical Cyber Security (RISCS)', thereby increasing awareness of the socio-technical security-thinking needed to acquaint operational technology actors in CNI sectors. This brings the benefits of considering both social and technical security risk factors while cooperating with security experts in the design of moderns CNI, especially those adopting emerging digital innovations.

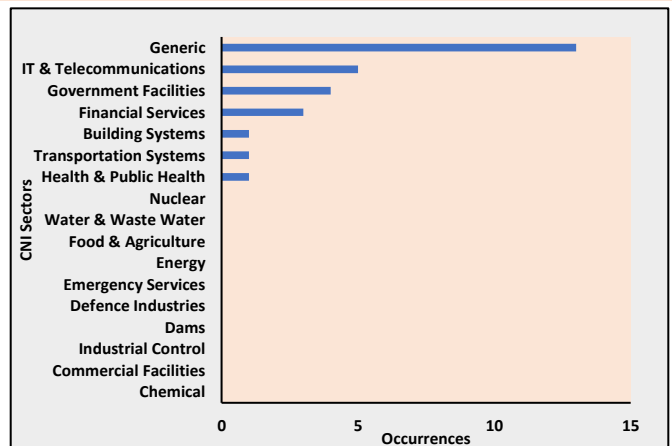

*Figure 2: Occurrence of Socio-Technical Applications in Infrastructure Sectors*

## 2. Promote the use of open-source M&S approaches to critical infrastructure security – tools and techniques

Some open-source simulators are developed for specific sectoral applications, and often for research and development purposes[18] as shown in Figure 3. This means that the benefits of these implementations are not necessarily shared as widely as possible. Although the active promotion of open source tools in security research and development is being progressed in the UK, there is still much to be done to grow the UK's stake in the global cybersecurity marketplace, especially in gaining traction on the use of open-source M&S for security and other purposes[19]. Driving policy in this direction could help lower the cost of security for a UK government that is trying to cope with shrinking budgets while managing the private ownership of CNIs. This applies more to the less security-minded CNI sectors such as the emergency services, food and agriculture[20].

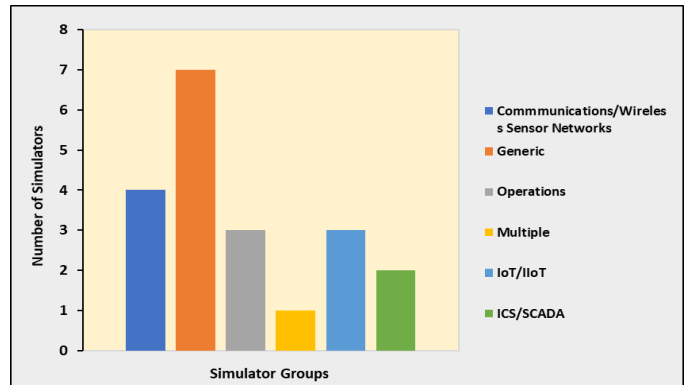

*Figure 3: Purpose-Driven Group Distribution for Open-Source Simulators*

*Government, through its departments and agencies can lead by example by establishing or revising procurement policies and funding contracts to encourage greater uptake of open-source M&S tools for CIP, especially where they perform comparably with commercial tools. This can drive the advantages of flexibility, reliability, reusability, simpler licence management, and best value for money.* Engaging the open-source security community can generate benefits beyond specific applications or sectors since gains can be made through establishing critical capabilities around better security and resilience. This can apply more broadly, which can also reduce the cost of owning and maintaining security tools under commercial arrangements.

## 3. Establish policies and platforms that encourage evaluating the credibility of security M&S approaches – tools and techniques.

The availability of data, mathematical expressions, algorithms, and prototype systems is currently inadequate to support a judgement of whether a security model or associated simulation is effective. To get the best security M&S outcomes, users must work closely with security analysts and developers throughout M&S creation, evaluation, and application. This is essential to establish confidence about what a security M&S tool or technique can and cannot do. Typically, three levels of evaluation for security M&S approaches/tools can help with this confidence-building, these include; *verification, validation,* and *accreditation* [21]. Verification answers: *Was the M&S tool built right?* Validation answers: *Was the right M&S tool built?* and Accreditation answers: *Is the built M&S tool believable enough to be used widely?*

Our research shows that a lot of industrial control system security M&S testbeds[1] do not undergo rigorous evaluation. A few go through verification and validation levels, but none appears to achieve any form of accreditation[6]. This could be due to the lack of standardised evaluation approaches and benchmarks, as well as weak corporate agreement and emphasis on the importance and strength of evaluation as a route to establishing M&S reliability and credibility.

Generally, evaluation can influence the design quality and credibility of M&S [22] by helping to demonstrate with appropriate evidence that modelled and simulated structures, setups, and associated outcomes are fit for their intended purpose, and do not present any intolerable risks. Having such evidence can support well-informed and confident decisions throughout a security M&S life cycle [23]. Impartial evaluation can increase confidence that appropriate sector/application-based guidelines for developing and using model security simulators and testbeds for research purposes have been followed. *Thus, it would be helpful to establish platforms/expert forums where the credibility of critical infrastructure security M&S tools (including testbeds) can be evaluated through independent reviews. This might be supported by policies that promote including/proving evidence of evaluation (e.g., verification, validation, and possibly accreditation/certification) for security M&S tools.* Public institutions, standardisation/certification bodies such as the National Physical Laboratory (NPL), British Standards Institution (BSI), the Institution of Engineering and Technology (IET), and the Centre for Digital Built Britain (CDBB) could help with this.

---

[1] *Test or experimental platform for executing activities and processes as if in a real environment.*

---

| 4. | **Create and maintain governance for security M&S** |

***A significantly beneficial influence can be achieved by implementing governance structures around the control and monitoring of guidelines for running and maintaining M&S projects funded or supported by the government.*** This might be similar to the Information Governance Management Framework for Nottinghamshire NHS Commissioning organisations[24], but more focused on security M&S processes, roles and responsibilities for development, quality and reliability assurance, and secure and ethical information sharing. Control and monitoring may be especially useful at the early stages of the security M&S projects related to new trends, such as IoT being incorporated into CNIs. This can enable a pathway for improving the availability, transparency and credibility of security M&S outcomes, open-source or otherwise.

| 5. | **Support open continuous cross-sector collaboration and knowledge exchange** |

***Policymakers should set up and drive policies that promote open continuous collaboration amongst research institutes and the private sector.*** Our research suggests that a larger proportion of known CIP approaches do not include any contexts that support the development of policies and regulations which might enhance security [9,10]. ***Driving policies that favour open cross-sector community collaboration, and continuous refinement of M&S approaches in response to evolving cybersecurity threats can lead to more relevant and accurate M&S and protection.*** For example, adopting a 4-dimensional (4D) modelling approach for security to help understand the complex time-dependent interactions of critical infrastructure systems with, and within built environments[25]. This can be useful theoretically and practically for visualising and analysing physical objects whose spatial relationships and security states change over time. The 4D approach can promote better expertise and knowledge exchange in the use and application of cutting-edge M&S for CNI protection. An expertise and practice base in this area can enable the UK to keep pace and be at the forefront of advanced security M&S technologies [26] as well as provide a platform for evaluating the effectiveness of M&S models, tools, and techniques via independent reviews. Such an expertise and practice base might include the involvement of NCSC, CPNI, Partners in the Construction Innovation Hub (BRE, MTC, CDBB), security and safety experts from industry (e.g., transport and water sector forums) and standards bodies like BSI or the IET, as well as cybersecurity researchers. Government may be better placed to encourage such partnerships due to its convening power, neutrality and regulatory roles.

| 6. | **Promote team-based collaborative development of security M&S tools amongst academia, industry and government.** |

***The government should encourage collaborative development of security M&S tools for critical infrastructure as opposed to relying on small, isolated teams of developers. This would improve the quality and future-proofing of security M&S.*** Collaborative development can bring the benefits of rigorous multiple independent testing and code reviews, which can reduce the operational risks of applying M&S while supporting generic solutions to common problems and encouraging innovation. The pace of change and the demand for innovative, efficient, and secure development of IoT and other emerging technologies means that where possible, good practice and insights should be reused or replicated within and across critical infrastructure sectors. This can help to avoid unnecessary duplication of investment and effort [19,27] and also drive the wider delivery of operational security, safety, and resilience.

| 7. | **Incentivise and/or reward the dissemination of security M&S research outcomes** |

Policies that incentivise the dissemination of security M&S research findings, together with implementation use cases, could catalyse strong and timely outcomes. For example, policies that encourage funding for critical infrastructure security M&S might benefit the wider economy and society in terms of strengthening security thinking and capability development [26]. ***It is recommended to create a policy and intervention strategy that encourages and rewards the adoption of high-quality security M&S as this can help keep the focus on safe, secure, and resilient outcomes.*** This would help to reduce barriers between critical infrastructure security research and application, and also encourage efforts to share research findings with wider user communities. Thereby, the costs of safety, security and resilience assurance could be reduced for other potential users in the same sector or in associated sectors.

## Conclusion

Before the era of digital network interconnectedness, most CNIs (while often physically interdependent) operated or functioned in silos. However, the integration of emerging technologies like the Internet, COTS devices, IoT and Edge processing, create further levels of connectivity and interdependency amongst infrastructure systems. These also mean that the cyber threat landscape of modern CNIs is widening with attacks increasing in frequency and sophistication with a greater probability of impact. Potentially adverse effects increasingly cross infrastructure boundaries; potentiating cascade effects and escalating failures through interdependencies.

Given the growing need to find ways of addressing the problem, M&S offers practical pathways, with techniques and tools to develop or improve security in the CNI domain. **Besides encouraging the development of technologies and methodologies, policymakers can also contribute in other ways including leading by example (e.g., in expert procurement, evaluation, and deployment), and by shaping and signposting the environment under which decisions and actions are initiated, in favour of M&S (in particular socio-technical) to support effective security.** Actions by policymakers can complement the work of industry and academia, and introduce improvements that enhance the responsiveness, use and utility of CNI M&S tools in this era of IoT and open-source code development. Such actions can translate into socio-economic impacts that could strategically drive more widespread adoption of security M&S approaches and tools, thereby establishing a culture of sharing lessons learned from the use and outcomes of security modelling for CNI protection.